\documentclass[twocolumn,prb,showpacs,preprintnumbers,amsmath,amssymb]{revtex4}
\usepackage{graphicx}

\usepackage{dcolumn}
\usepackage{bm}

\usepackage{verbatim}   
\newcommand{\br}{\mathbf{r}}

\newcommand{\bn}{\begin{equation}}
\newcommand{\ee}{\end{equation}}
\newcommand{\bga}{\begin{eqnarray}}
\newcommand{\eda}{\end{eqnarray}}
\newcommand{\half}{\frac{1}{2}}
\newcommand{\diff}{\mathrm{d}}

\newcommand{\R}{R30$^\circ$}

\newcommand{\chalmersMC}{$^{1,*}$Department of Microtechnology and Nanoscience, MC2, Chalmers University of Technology,
SE-41296 G\"{o}teborg, Sweden}
\newcommand{\maryPhysics}{$^{2,\dagger}$Department of Physics, University of Maryland, College Park, Maryland 20742-4111}

\begin{document}

\title{Rings sliding on a honeycomb network:
Adsorption contours, interactions, and assembly of
benzene on Cu(111)}

\author{K. Berland$^{1,*}$, T.L. Einstein$^{2,\dagger}$, and P. Hyldgaard$^{1,*}$}
\email[]{berland@chalmers.se}
\email[$^\dagger$]{einstein@umd.edu}
\email[$^{**}$]{hyldgaar@chalmers.se}
\affiliation{\chalmersMC \\ \maryPhysics}

\date{\today}

\begin{abstract}
Using a van der Waals density functional (vdW-DF) [Phys.~Rev.~Lett. {\bf 92}, 246401 (2004)],
we perform {\it ab initio\/} calculations for the adsorption energy of
benzene (Bz) on Cu(111) as a function of lateral position and height.  We find that the vdW-DF inclusion of nonlocal correlations (responsible
for dispersive interactions) changes the relative stability of eight
binding-position options and increases the binding energy by over an
order of magnitude, achieving good agreement with experiment.  The admolecules can move almost freely along a honeycomb web of ``corridors" passing between fcc and hcp hollow sites via bridge sites.  Our diffusion barriers (for dilute and two condensed adsorbate phases) are consistent
with experimental observations.  Further vdW-DF calculations suggest that the more compact (hexagonal) Bz-overlayer phase, with lattice constant $a\! =\! \hbox{6.74 {\AA}}$, is due to
direct Bz-Bz vdW attraction, which extends to $\sim\hbox{8 \AA}$.  We attribute the second, sparser hexagonal Bz phase, with $a\! =\! \hbox{10.24 {\AA}}$, to indirect electronic interactions mediated
by the metallic surface state on Cu(111).  To support this claim, we use a formal Harris-functional approach to evaluate nonperturbationally the asymptotic form of this indirect
interaction. Thus, we can account well for benzene self-organization on Cu(111).
\end{abstract}
\pacs{68.43.-h,73.20.-r,71.15.Mb,73.90.+f}
\maketitle
\section{Introduction}


With adsorption of atoms and of simple molecules on metals relatively well understood,
interest has turned to the adsorption of organic molecules.\cite{Salaneck:Book}
There are many technological motivations for studying adsorption of electronically
functional organic overlayers,  {\it e.g.,} organic light-emitting diodes (OLEDs)
offer great promise for displays.\cite{Hung:MSE}  Acenes and acene derivatives
are of particular interest, not just because of their importance in organic
semiconductors but also because of the self-organized structures that they form on surfaces.

Arguably the most dramatic example of the latter is the giant regular honeycomb network
formed by anthraquinone (AQ) on Cu(111).\cite{Bartels:AQ}  The pore diameter is
unprecedentedly large, over 5nm, and each cell encloses over 200 uncovered Cu surface
atoms.  Our understanding is that the adsorption process is primarily due to van der
Waals (vdW) interactions, the short-range arrangement is due to hydrogen bonding from
an aromatic hydrogen on one AQ to an oxygen of a carbonyl group on its neighbor;
we believe that the long-range regularity can be attributed to indirect interactions
mediated by the metallic (Shockley) surface state centered around the zone center
of noble metal (111) surfaces,\cite{Kevan:Surf} in a mechanism analogous to that
documented for atomic adsorbates.\cite{exp:Surf, theo:Surf1, theo:Surf2, theo:Surf3, NEWREF, bfDFT:Surf}

In a first attempt to understand the details of the adsorption
and self-organization of acene derivatives, we here consider the
simplest member of this family, the monocyclic aromatic ring benzene (Bz).
Adsorption of Bz on Cu(111) displays remarkable molecular-adsorption dynamics with
observations\cite{Stevens:exp, Munakata:exp, freely-Science:Weiss, freely-Surface:Weiss, Ertl:exp, Woll:exp, Dan:exp}
of high  mobility of individual Bz molecules at low temperatures,
as well as a possibility for the Bz adsorbates to condense into two coexisting
adsorbate-overlayer phases, C1 and C2, characterized by hexagonal nets with lattice
constants \hbox{10.24 \AA} and \hbox{6.74 \AA}, respectively.
For Bz, as for AQ, the adsorption energy arises primarily from
van der Waals (vdW) interactions with conduction electrons on the Cu
surface, with very small orbital changes and substrate-adsorbate
charge transfers.\cite{Lomas}

The case of Bz on Cu(111), Fig.~1, provides a forum at which to show the power of
a recently-developed vdW density functional (vdW-DF)\cite{Dion:vdW, Timo:vdW, Review:vdW}
to describe molecular physisorption on a system with extremely low corrugation.
Systems with low corrugation are interesting and challenging since they leave
the adsorbates free to respond to weak direct and indirect mutual interactions
which must therefore be described and analyzed in detail. Traditional density
functional theory (DFT) calculations, based on local or semi-local approximations
for the functional, provide detailed information about the variations in
atomic-adsorbate energies\cite{diffCu:Stoltze,Cu111SiteDeterm,diffCu:Marincia} but
cannot account for dispersive interactions.\cite{Review:vdW, HardNumbers, layer:vdW}
Thus, traditional DFT calculations have only a limited ability to compare
binding energies, diffusion and rotation barriers for the molecular adsorption
and does not permit an analysis of the adsorbate self-organization.
In particular, the generalized gradient approximation (GGA) has proved inadequate
for Bz on Cu(111).\cite{Ford:BenzeneCuOld}  In this paper we report vdW-DF
calculations of the subtle variation in the adsorption energy of individual molecules
with bonding site, of the intermolecular vdW binding, and of the energy barriers
for adsorbate  dynamics at different adsorbate condensations.

\begin{figure}[ht]
\includegraphics[width=8cm]{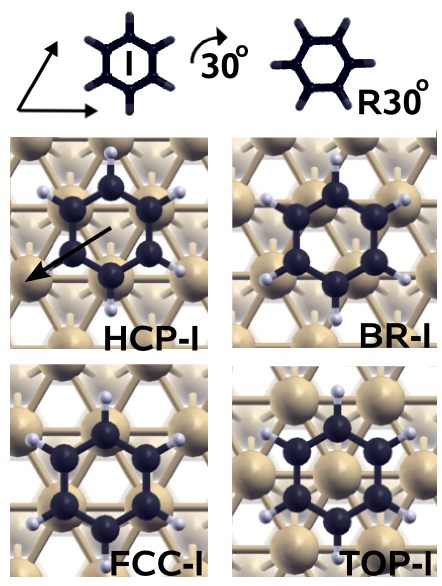}
\caption{[Color online] Schematics of the eight high-symmetry configurations of adsorbed
benzene (Bz) on Cu(111). High-symmetry configurations are found when the center of the molecules
align with the traditional characteristic atomic-adsorption sites of the Cu(111) surface
(four main panels).  In the illustrated I (``in-phase") orientation, one of the two equilateral triangles formed by alternate C's of the ring sits above three Cu atoms in the surface for HCP and FCC; for BR and TOP, these 3 carbons can be similarly aligned via a translation.  Orientation R30$^\circ$ differs from orientation I
by a 30$^\circ$ rotation of the Bz adsorbate, as shown in the top pair of sketches.  The arrow in upper left main panel illustrates
a possible diffusion path that connects the high-symmetry sites, running from an HCP-I site,
over a bridge site BR-I, then over an FCC-I site, and finally to a TOP-I site.  A similar path could occur for the \R orientation.
}
\label{fig:CuLoc}
\end{figure}

Furthermore, we use the vdW-DF results to analyze and possibly
interpret the self-organization mechanisms observed for Bz adsorption
on Cu(111).\cite{Dan:exp} Assuming that the admolecules form a close-packed
(hexagonal) structure, we determine its lattice constant from the minimum
of the interaction potential.  {\it For the more compact C2 phase,} our
vdW-DF calculations of the direct vdW binding indicate a preference for
a lattice constant of \hbox{6.9 \AA}.  Given the tendency of the vdW-DF method
to slightly overestimate binding separations,\cite{Review:vdW, IJQC:vdW} this result
suggests that the C2 phase can be interpreted as a direct, vdW-bound molecular-lattice
structure.  Explicit vdW-DF predictions of the preferred Bz orientations and of
rotational barriers may offer experimental tests of this interpretation.
{\it For the less dense C1 phase} we find that the lattice spacing lies beyond the
range of substantial bonding via the direct intermolecular vdW binding. However,
use a Harris-functional scheme\cite{Harris, theo:Surf3} formally extends our DFT
analysis and shows that the C1 lattice spacing agrees excellently with the first energy
minimum of the indirect electronic interaction\cite{theo:Surf1, theo:Surf2, theo:Surf3} mediated by
the Cu(111) metallic surface-state (Shockley) band. Since we have explicitly verified
that the Cu(111) surface state survives the addition of Bz molecules, we find it
plausible that this indirect electronic interaction underpins the self-organization in
the C1 phase.

This paper has the following plan: in Sec.~II we summarize the vdW-DF theory
and formal Harris-functional evaluation that we use, along with various computational details.
In Sec.~III we present our results for adsorption energies, diffusion barriers, and barriers
to rotation.  In the next section we discuss these results, emphasizing possibilities for
crystal organization.  The final section contains our conclusions.

\section{Density Functional Theory}

A recent vdW density functional, vdW-DF,\cite{Dion:vdW, Timo:vdW} has proved useful
in describing a variety of sparse matter,\cite{Review:vdW} {\it i.e.\/} systems which contain
important regions of low electron density where dispersive interactions contribute
significantly to the materials' coherence.  The vdW-DF method provides a transferable account
of general materials and bindings, ranging, e.g., from the all-covalent C-C bonding within a graphene
sheet to the interlayer dispersive bonding in graphite and graphite intercalates.\cite{Eleni:vdW}
DFT calculations using the vdW-DF has successfully characterized the adsorption
of Bz on the semi-metal graphene\cite{Svetla:BenzeneGraphite} and on the semiconductors
silicon\cite{Karen:BenzeneSilicon} and MoS$_2$.\cite{Review:vdW} A vdW-DF study also documents
that vdW interactions (nonlocal correlations) significantly affect the anchoring of phenol on
an oxide.\cite{Svetla:phenol} There are fewer applications for the interaction between
organic molecules and metal surfaces, but very recent examples include the adsorption on Cu(110)
of triophene by Sony {et al.}\cite{Sony:Cu110} and of
Bz and other organic molecules by Atodiresei {et al.}\cite{Blugel:Cu110}

We describe the physisorption and two-dimensional assembly of Bz on Cu(111) within DFT by combining
large-scale vdW-DF calculations and a formal Harris-functional evaluation,\cite{theo:Surf3} for:
(a) adsorption of isolated Bz molecules and (b) direct and indirect (substrate-mediated)
interactions between Bz adsorbates.  Brute-force vdW-DF calculations treating the physisorption
and the inter-adsorbate interactions in one giant unit cell are not computationally tractable for
Bz adsorbates on Cu(111).  Instead, we apply large-scale vdW-DF calculations to characterize the molecular
adsorption both in the case of a frozen surface and in the presence of adsorption-induced surface
relaxations.  A separate set of vdW-DF calculations investigate the direct molecular interaction between
pairs and in a regular three-dimensional lattice by ignoring the coupling to the substrate and by focusing
strictly on the two-dimensional arrangement of adsorbates. Finally, we use a general, analytical
Harris-functional evaluation\cite{Harris, theo:Surf3} for a nonperturbative treatment of the indirect electronic interaction between
adsorbates to formally complete an approximative DFT characterization of the adsorbate interaction
and of the two-dimensional self-organization of Bz on Cu(111). This indirect
electronic interaction is oscillatory and very long range; it arises from interference of Friedel
oscillations produced by the strong adsorbate-induced scattering of surface-state
electrons.\cite{eigler,exp:Surf,theo:Surf3} For atomic adsorbates, the indirect
electronic interaction can be described in traditional DFT implementations using
very large scale unit cells;\cite{bfDFT:Surf} the present vdW-DF study must, in
practice, rely on an implicit parameterization given by the detailed nature of the
adsorbate-induced scattering.\cite{theo:Surf2, theo:Surf3,Olsson}

Our splitting of the physisorption and assembly problems permits us to provide explicit
vdW-DF determinations for some properties but can, naturally, only serve as an approximate
characterization of other properties.  Our description of the combined interactions of
the adsorbates ignores, for example, the contributions to the direct vdW binding
between molecules from small, adsorption-induced changes in the electron density distribution.
Such effects should be present even if our use of a Bader analysis\cite{Bader:main,Bader:alg} documents
that the total adsorbate-to-substrate charge transfer vanishes. Also, the precise nature of the surface-state
scattering induced by Bz adsorption is not known. Lacking a detailed experimental characterization,
we can only apply our formal Harris-functional evaluation to provide a qualitative description of the
long-range indirect electronic interaction\cite{exp:Surf,theo:Surf1,theo:Surf2, theo:Surf3} between adsorbates.
Nevertheless, we argue that the main qualitative results, possible interpretations of the C1 and C2
phases, are robust.  Moreover, the separation of the general physisorption and assembly problems should not
significantly affect the results for the dilute-phase adsorption energies, the minute surface
corrugation, and the enhancement of the C2 formation energy by direct Bz-Bz vdW bonding.

\subsection{van der Waals density functional calculations}

Since details of the vdW-DF method and of the functional design are given in
Refs.~\onlinecite{IJQC:vdW,Dion:vdW,Timo:vdW, Review:vdW}, we here only summarize the
calculational approach.  For every atomic configuration in a unit cell $V_0$, we obtain
the self-consistent charge with traditional DFT, typically as parameterized in PBE-GGA.\cite{PBE}
From the known electron-density variation, we also calculate a local wavevector $q_0(\mathbf{r})$
which characterizes the energy density in a GGA.\cite{Timo:vdW}  Next, in a post-processing phase,
we calculate a nonlocal-correlation term
\cite{Dion:vdW}
\bn
E_c^{\mathrm{nl}}[n] = \frac{1}{2} \int_{V} d\mathbf{r} \int_{V_0} d\mathbf{r}'
n(\mathbf{r})
\phi(\mathbf{r},\mathbf{r}')
n(\mathbf{r}')
\label{eq:Ecnl}
\ee
as a spatial double integration over an interaction kernel $\phi(\mathbf{r},\mathbf{r}')$, where the latter
integration is only over the unit cell $V_0$.  The kernel can be conveniently evaluated from a tabulated, universal
form by introducing the two dimensionless variables $\Xi \equiv (q_0+q_0')|\mathbf{r}-\mathbf{r}'|$ and
$\xi \equiv (q_0-q_0')/(q_0+q_0')$.  The nonlocal correlation term $E_c^{\mathrm{nl}}[n]$ describes the
interactions mediated via coupling to the electrodynamical field and allows vdW-DF to describe sparse-matter
behavior. We also evaluate a (semi-local) part $E_0$ which is obtained from the PBE-GGA energy
by adjusting the exchange to the revPBE\cite{revPBE} form and the correlation
to the value in in the local-density approximation.\cite{vwn} This term retains information about, for example,
the kinetic-energy repulsion that adds to the surface corrugation.\cite{Nanotube:vdW}
Finally, we determine the total vdW-DF energy as the sum
\bn
E^{\mathrm{vdWDF}}[n]=E_0[n]+E_c^{\mathrm{nl}}[n]
\ee
in a scheme that is transferable and capable of describing interactions for both high and low
electron concentrations.\cite{IJQC:vdW, Review:vdW}

In many respect the present vdW-DF calculations are similar to those presented in
Refs.~\onlinecite{Svetla:phenol,Eleni:vdW,Svetla:BenzeneGraphite,Karen:BenzeneSilicon,Jesper:vdW,Nanotube:vdW}.
The physisorption system requires us to resolve very small energy differences (e.g., the variation in adsorption energy with site) and to
ensure accurate and well-converged evaluations of changes in the nonlocal correlation $E_c^{\mathrm{nl}}[n]$.
Since Eq.~(\ref{eq:Ecnl}) involves a real-space integral, the evaluation of $E_c^{\mathrm{nl}}[n]$ depends slightly on the grid-sampling because of the rapid variation
in the kernel at small $\Xi$.\cite{Dion:vdW,Nanotube:vdW}  We employ a number of additional calculational steps to enhance accuracy and test the convergence in the evaluation of Eq.~(\ref{eq:Ecnl}):

We characterize a general
adsorbate system `$A$' by the Bz configuration $a$ and the associated substrate configuration $S_a$
(which may be the frozen free surface or the surface specified by adsorbate-induced relaxations as obtained
by calculations and implementations of GGA-PBE forces).  For a given atomic configuration $A$, the
nonlocal correlation energy~(\ref{eq:Ecnl}) will be dominated by contributions from core regions at
or close to the atomic positions; this causes a very small relative error $\delta E_c^{\mathrm nl}(A)$
that varies when atoms are displaced relative to the density grid.  To compute adsorption energies and
compare two systems, $A$ and $B$, the relevant nonlocal-correlation energy difference is instead specified
by the smaller $E_c^{\mathrm nl}$ contributions that arise outside the atomic cores. However, since the
two systems have different atomic configurations (and hence different alignment with the density grid),
the small relative errors [$\delta E_c^{\mathrm nl}(A)$ and $\delta E_c^{\mathrm nl}(B)$] need not
always cancel and could affect the the accuracy for systems with minute energy differences. In this study
we therefore avoid direct evaluations of the nonlocal-correlation energy differences $E_c^\mathrm{nl}(A)
- E_c^\mathrm{nl}(B)$ and apply, instead, the procedure described in
Refs.~\onlinecite{Svetla:phenol, Nanotube:vdW}.  In practice, we determine
$E_c^\mathrm{nl}(A) - E_c^\mathrm{nl}(B)$ in a sequence of steps where we separately calculate
the $E_c^\mathrm{nl}$ changes associated with removing the Bz in configuration $a$ ($b$)
to infinity, always keeping the atomic positions frozen and fixed relative to the density grid.
In the vdW-DF study of the effects of adsorbate-induced relaxations of the substrate, we follow
these steps by a GGA-PBE calculation of the energy difference between $S_a$ and $S_b$.  A previous
study\cite{Eleni:vdW} documents that a grid spacing of 0.15 {\AA} is sufficient to achieve
convergence of the vdW-DF binding energy in adsorption problems when combined with the above-mentioned scheme.

The calculations of $E_c^{\mathrm{nl}}[n]$ via Eq.~(\ref{eq:Ecnl}) are accelerated by introducing a cut-off radius $R_{\mathrm{cut}}$ while
carefully testing the convergence of the spatial double integration.  Contributions of the long-range dispersive interactions
to $E_c^{\mathrm{nl}}[n]$ fall off slower with distance $|\mathbf{r}-\mathbf{r}'|$ for the
adsorbate-slab  interactions than for the Bz-Bz interactions. The $\mathbf{r}'$ integration
in Eq.~(\ref{eq:Ecnl}) exceeds the unit cell but the integrated weight of contributions outside
$R_{\mathrm{ cut}}$ still decreases fairly rapidly with the value of
$R_{\mathrm{cut}}$.\cite{Svetla:BenzeneGraphite, Jesper:vdW}  We have explicitly
tested that the choice $R_{\mathrm{cut}}= \hbox{23 \AA}$ ensures that
$E_c^{\rm nl}$ converges at the sub-meV level.

\subsection{Long-range indirect electronic interactions between adsorbates: nonperturbative $s$-wave treatment in the asymptotic regime}

To complete our DFT analysis we must explicitly add consistent estimates for
the indirect, substrate-mediated interactions.  Benzene adsorbates not only
interact directly through vdW forces, but also through forces mediated by
the Cu(111) substrate.  Elastic-deformation effects,\cite{Elastic} dipolar
couplings,\cite{KL76} and indirect electronic interactions\cite{theo:Surf1,theo:Surf3} all
exhibit a power-law decay \cite{theo:Surf2,theo:JKN} and can affect the adsorbate binding at distances $d> 8$ {\AA}
where (we find) the direct adsorbate-adsorbate vdW interaction is no longer substantial.
Since we calculate the direct Bz-Bz vdW interaction in a separate vdW-DF study
that ignores the substrate, we must separately analyze the set of indirect interactions
and add the relevant DFT-based terms.

\begin{figure}[ht]
\centering
\includegraphics[width=9cm]{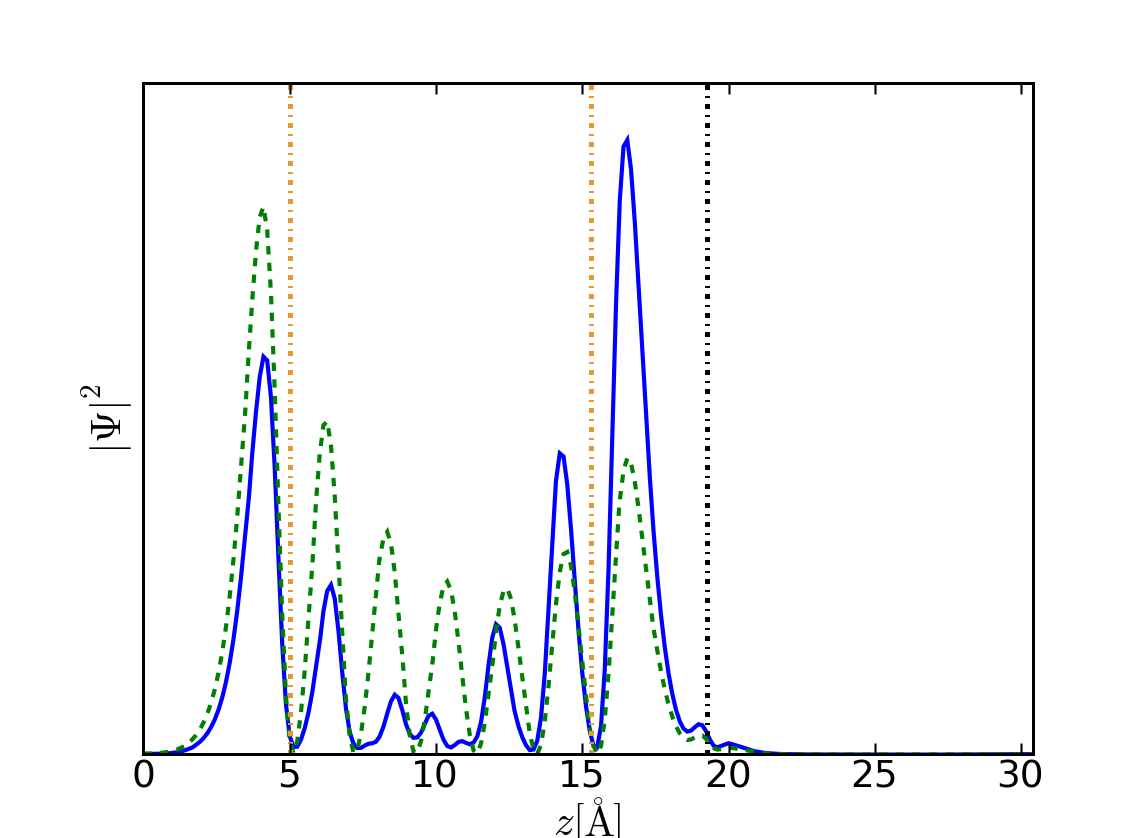}
\caption{Robustness of Cu(111) surface-state formation to Bz adsorption and evidence of
adsorbate-induced surface-state scattering presented in the dilute limit with one Bz molecule on
a frozen 3$\times$3 Cu(111) supercell. The figure shows the variation of the in-plane average of the surface-state wave functions squared, $|\Psi|^2= A^{-1} \int_{\rm 2D cell} \diff A \, |\psi_{\rm surf-st.}|^2$ (evaluated at the $\bar{\Gamma}$ point) along the
surface-normal coordinate $z$ for our choice of a 6-layer copper slab (with top and
bottom Cu layers identified by light dashed lines). The physisorption position of the Bz adsorbates
is indicated by the black dashed line (near $z=19\mathrm{\AA}$).  Many layers are necessary to fully describe the surface-state
properties and, in a slab geometry, to decouple the surface state on the top and on bottom sides. Our choice of
6 copper layers proves to be the smallest that exhibit clear signs of a Shockley surface state
--- an exponential decay of the surface state into the bulk and a large spill-out far beyond the
outer layers.  The approximative symmetry of the top-side (solid curve) and bottom-side (dashed curve)
surface states implies that the surface states survive the physisorption in the dilute limit. At the same
time, the oscillatory behavior evident at the position of the adsorbate
shows that the Bz interacts with surface-state electrons.}
\label{fig:SurfSurv}
\end{figure}

Of particular interest is the surface-state-mediated indirect interaction.
This metallic surface state is responsible for the dramatic images obtained
with scanning-tunneling microscopy (STM).\cite{eigler,exp:Surf}  Fig.~\ref{fig:SurfSurv}
shows that there is notable coupling between the Bz adsorbate and the surface-state
wavefunction even in a dilute-coverage regime.  This conclusion is expected from the analogous case
of atomic adsorption\cite{Olsson} on Cu(111)\cite{exp:Surf} where the measured adsorption-induced phase shifts
are known to be large -- $\Theta=\Theta_F\sim \pi/2$ --- indicating that a perturbative treatment
of the surface-state-mediated adsorbate-pair interaction is inadequate.  Our vdW-DF calculations of
Bz physisorption suggest that the dipolar and elastic interactions between Bz adsorbates should be
insignificant: a Bader analysis\cite{Bader:main,Bader:alg} documents vanishing electron transfer
and we find only very small adsorbate-induced relaxations, $\sim\hbox{30 meV}$, which do not
change the relative stability of the various sites. However, Fig.~\ref{fig:SurfSurv} shows that
the metallic surface state survives the Bz adsorption and therefore can mediate an interaction
between adsorbates. Fig.~\ref{fig:SurfSurv} also demonstrates that Bz affects the surface-state scattering and produces, in turn,
a long-range indirect electronic interaction.\cite{exp:Surf,theo:Surf1,theo:Surf2,theo:Surf3}
The minor adsorbate-induced relaxations will only enhance the surface-state scattering
by increasing the coupling between the surface-state and evanescent bulk-like
states around the adsorption site.

Unlike the other adsorbate-adsorbate interactions, the indirect electronic interaction is oscillatory. At short range, it
depends on all the occupied states but asymptotically the behavior is dominated by the \textbf{q}-state at the
Fermi level whose velocity is parallel to the separation vector \textbf{d} between the adsorbates.\cite{theo:Surf2}
While this interaction is typically anisotropic, the metallic (Shockley) surface-state band on Cu(111) is
circularly symmetric, so the interaction becomes isotropic, and the relevant \textbf{q$_F$} is parallel to \textbf{d}. These conditions lead to the perturbative treatment well known as the Ruderman-Kittel-Kasuya-Yosida (RKKY) interaction,\cite{RKKY,kittelrev} which has recently been applied to actively-investigated low-dimensional systems.\cite{RKKYloD}
A significant effect of mediation by metallic surface states is that the envelope decays
exceptionally slowly, $\propto d^{-2}$ rather than the usual $\propto d^{-5}$ for mediation
by bulk states at metal surfaces.\cite{exp:Surf,theo:Surf1,theo:Surf2,theo:Surf3}

The primary modification due to a nonperturbative treatment is the addition of a phase shift in the argument of the sinusoidal function in the RKKY expression (and for Friedel oscillations in general).  To explicate this result, we extend our regular vdW-DF calculations with a
Harris-functional scheme,\cite{Harris} which formally constitutes an
asymptotic DFT evaluation\cite{theo:Surf3} of the indirect electronic interactions
between Bz adsorbates. This choice permits a consistent description of the nonperturbative
scattering and interaction effects arising with general (complex) s-wave surface-state phase
shifts.\cite{theo:Surf2} The complex phase shift reflects transitions both among surface-state wavevectors and into evanescent bulk states.
A simple s-wave characterization,\cite{theo:Surf3} with simultaneous descriptions of both
scattering channels, has proven reasonably successful in estimating the resulting indirect-electronic
interaction for a pair of atomic adsorbates\cite{exp:Surf}; the s-wave characterization should also be
applicable for Bz on Cu(111) because the coupling of Bz to the substrate occurs in a relatively confined area.

\begin{figure}[ht]
\centering
\includegraphics[width=8cm]{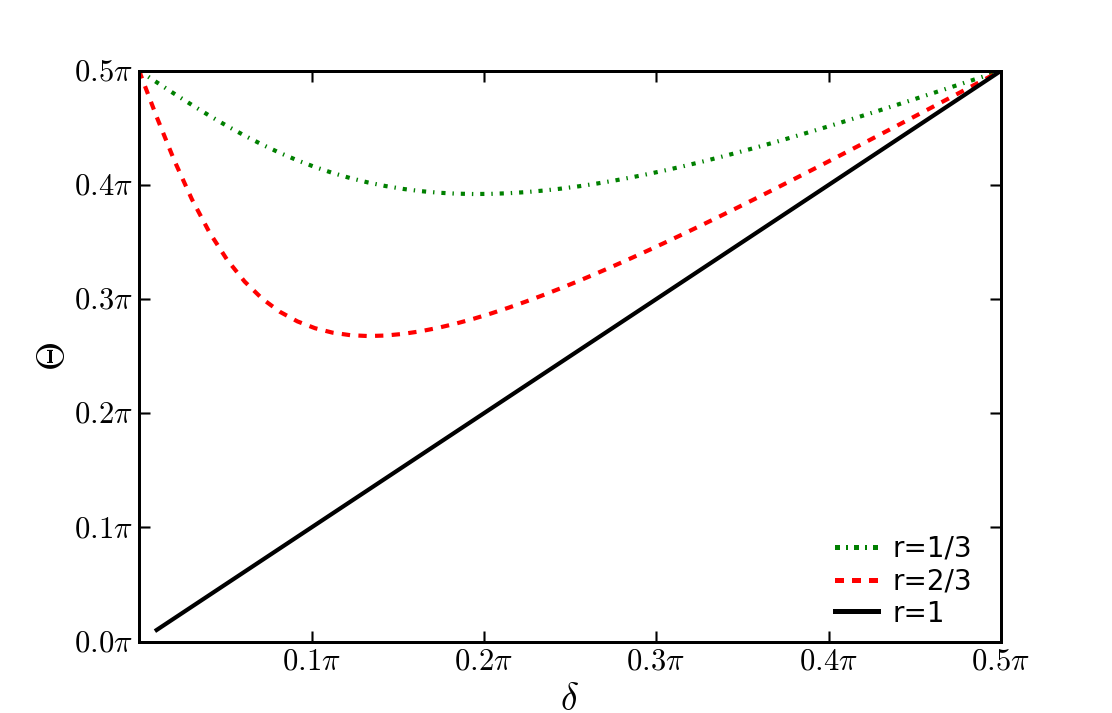}
\includegraphics[width=8cm]{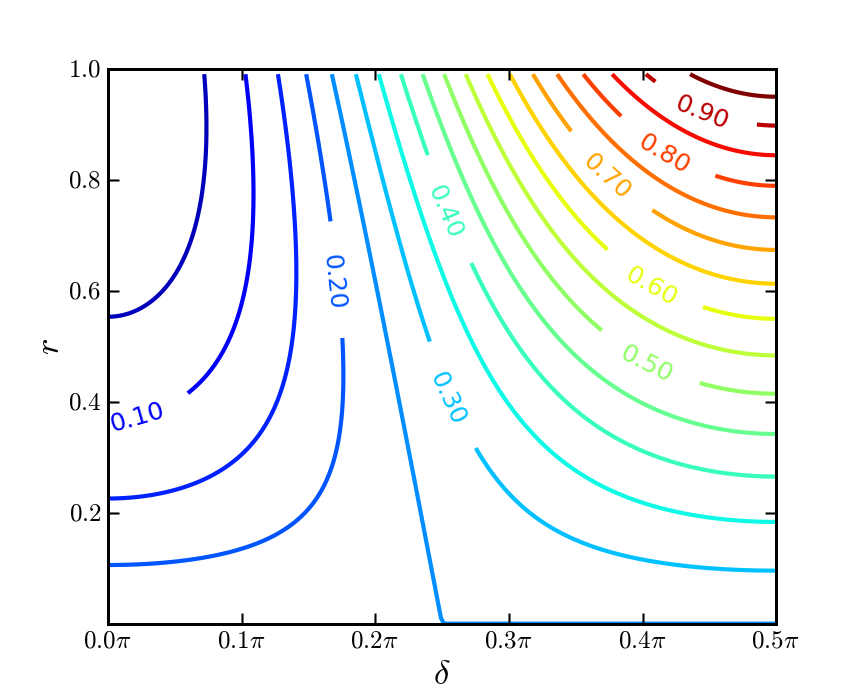}
\caption{[Color online] Dependence of effective interaction phase shift $\Theta$ (top panel) and scaled interaction strength $A$
of indirect electronic adsorbate-pair interaction [Eq.~(\ref{eq:pair})] with detailed nature of adsorbate-induced
s-wave scattering of surface-state electrons. The variation is expressed by the phase-shift for
intra-surface-state transitions $\delta$ and the intra-surface-state reflectance $r$ (where $1-r$ describe the
probability that the adsorbate causes a scattering from a surface state into an evanescent bulk-like state).
Analogy with the case of interactions between Cu(111) adatoms suggests that it is reasonable to expect
both $1-r>0$ and $\delta > 0$. While we do not have explicit determinations of the relevant parameters for
Bz on Cu(111), the panels suggest that values, $\Theta > \pi/3$ and $A > 0.1$, can be considered typical
for indirect electronic adsorbate pair interactions [valid for the present case of Bz on Cu(111)
as well as for Cu adatoms on Cu(111)\cite{exp:Surf}].}

\label{fig:IndirElecParamSpace}
\end{figure}

The s-wave formulation defines an effective Fermi-level T-matrix $t_0(\epsilon_F)\propto [\exp(i2\delta_0)-1]$
given by a complex phase-shift $\delta_0=\delta+i\delta''$ and formally expressing the probability
for transition entirely within the surface-state channel. Unlike for a strictly 2D scattering system, however,
the intra-surface-state (ISS) scattering is given both an ISS phase shift $\delta$ and an ISS
reflectance $r=\exp(-2\delta'')$. The latter corresponds to a bulk-state absorption probability $1-r$, which
cannot vanish since the adsorbates cause surface-state electrons to scatter into bulk states.
The nonperturbative analysis leads to an effective Fermi-level phase shift
$\Theta$ which can be obtained from\cite{theo:Surf3}
\begin{eqnarray}
\lefteqn{\tan[\Theta(r,\delta=\delta_F)] = \frac{1-r\cos(2\delta)}{r\sin(2\delta)}
\quad \Rightarrow} \\
 \Theta & \! \sim\!& \frac{\pi}{2} \! -\! \frac{2r}{1 \! \mp \! r}\varphi
+ \frac{4(r \! \pm \! r^2)}{3(1 \! \mp \! r)^3}\varphi^3  \! +\! {\cal O}\left( \varphi^5\right) \nonumber
\label{eq:thetadef}
\end{eqnarray}
\noindent where the expansion is around either endpoint 0 or $\pi/2$, with $\varphi = \delta$ or $(\pi/2)  \! -\! \delta$, with the upper or lower sign applying, respectively.  For small $r$, $\Theta(r,\delta)$ is nearly symmetric about $\delta = \pi/4$, descending linearly (from value $\pi/2$), with a shallow slope, from either endpoint.  As $r$ increases, the magnitudes of the linear slopes increase asymmetrically, with a shift of the minimum $\delta_{\rm min}$ to progressively smaller values, while the minimum value of $\Theta$, $\Theta(r,\delta_{\rm min})$, decreases continuously from $\pi/2$ to 0; explicitly,
\bn
\delta_{\rm min} = \frac{1}{2} \cos^{-1}r, \quad \Theta(r,\delta_{\rm min}) = \tan^{-1}\frac{\sqrt{1-r^2}}{r}.
\label{eq:min}
\ee
\noindent The upper panel of Fig.~\ref{fig:IndirElecParamSpace} depicts
the relation between $\Theta$ and $\delta$ for three representative $r$ values.

We determine $E^{\mathrm{ans}}_{\mathrm{pair}}(d)$, where the superscript denotes that this
is the {\it a}symptotic form of the indirect-electronic adsorbate-pair interactions,
computed with the {\it n}onperturbative formal Harris-functional for {\it s}-wave
scattering, through the analytical evaluation\cite{exp:Surf,theo:Surf2}
\bn
E^{\mathrm{ans}}_{\mathrm{pair}}(d)
= - A(r,\delta) \left(\frac{4 \epsilon_F}{\pi^2}\right)\frac{\sin(2 q_Fd +2\Theta( r,\delta))}{(q_Fd)^2}.
\label{eq:pair}
\ee
Here
\bn
A(r,\delta) = \left[
(1-r)^2/4
+
r\sin^2(2\delta)
\right]
\label{eq:Adef}
\ee
represents a scaled, dimensionless interaction strength, $0\leq A\leq 1$. For Cu(111) the wavevector $q_F$ of the
surface state is 0.22 $\mathrm{\AA}^{-1}$, and the Fermi energy $\epsilon_F$ is 0.39 eV. \cite{Kevan:Surf}
The explicit values of $\delta$ and of $r$, or equivalently $\Theta$ and $A$, depend on the details of
the Bz-induced scattering, for which we have no experimental characterization; we stress that, in the
present s-wave analysis, the effective interaction phase is the same $\Theta$ that characterizes the STM measurements
of the local density-of-state variation.\cite{eigler,exp:Surf}

Fig.~\ref{fig:IndirElecParamSpace} displays the possible variation in interaction phase
$\Theta$ (upper panel) and scaled interaction strength $A$ (lower panel) with the s-wave modeling
parameters $\delta$ and $r$.  The perturbative regime of RKKY corresponds to the
limit $(r \to 1,\delta\to 0)$, while $r \to 0$ corresponds to the case of completely absorbing
scattering (all surface-state electrons scatter into the bulk), for which Eqs.~(\ref{eq:thetadef})
and (\ref{eq:Adef}) indicate $\Theta=\pi/2$ and $A=0.25$, respectively. In making these plots we
have implicitly assumed a positive interaction phase shift $\Theta>0$ (as was measured for atomic
adsorbates\cite{exp:Surf}). We expect Bz adsorbates to produce significant coupling between surface
and bulk states, $r \neq 1$ and finite ISS scattering $\delta \neq 0 $.
The lower
panel of Fig.~\ref{fig:IndirElecParamSpace} shows a contour plot of the scaled interaction strength $A$; even moderate values of
the ISS scattering phase-shift $\delta$ and of the bulk-state absorption, $1-r$, produces
significant values of the interacting strength. Generic values for the ISS scattering parameters
($\delta$ and $r$) produce an interaction strength and phase which are comparable to or larger
than those characterizing interactions between Cu adatoms on Cu(111):\cite{exp:Surf}
$A_\mathrm{Cu/Cu}=0.08$ and $\Theta_\mathrm{Cu/Cu}=\pi/3$.

It is interesting to compare the indirect electronic interactions between pairs of
Bz and pairs of Cu adatoms on Cu(111). In both cases is there is essentially no charge transfer.
The Friedel sum rule\cite{FriedelSum} then tells us\cite{theo:Surf3} that the coupling between surface and
bulk-like states must be the primary cause of the Friedel oscillations. On
the one hand, since Bz is physisorbed to Cu(111), its impact should be smaller than chemisorbed Cu adatoms,
with a smaller adsorbate-induced coupling between the surface and bulk states.\cite{eigler,exp:Surf} On
the other hand, Bz has a quadrupole moment, defined by a total of 6 C-H bonds with a nonzero dipolar moment
that will scatter the surface-state electrons. In a simple tight-binding model of weak adsorption, the indirect interaction varies as the square of the adsorption energy.\cite{theo:Surf2}  It is plausible that scattering off each carbon atom in Bz
produces an indirect electronic pair interaction with a nonvanishing interacting strength and phase, so
that there is a significant (though likely not 36-fold) enhancement of the individual
interactions.  One might instead view the adsorption as due to a molecular orbital of each cyclic group, implying that the indirect interactions of anthracene pairs might be 3 - 9 times those of benzene at the same separation. Lukas et al.\cite{Woll:exp} did find the intriguing result that the binding energies of C$_{\rm 4n+2}$H$_{\rm 2n+4}$ on Cu(111) for n=0,1,2,3 (acetylene, benzene, naphthalene, anthracene) are approximately collinear as a function of $n$, with the binding energy of anthracene about 5 times that of acetylene and somewhat more than twice that of benzene.  In summary, it is not clear how several adsorbates within the screening length of the bulk electrons
couple to the surface state compared to just a single adatom.  This issue begs further investigation.

\subsection{Adsorption and surface assembly energies}

The adsorption energy is given by the difference between the total energy for a given configuration and
the total energy of the Bz and the surface in isolation
\bn
E^{\rm vdWDF}_{\rm ads} = E^{\rm vdWDF}(h)-E^{\rm vdWDF}(h\rightarrow \infty) \label{eq:ads}.
\ee
\noindent (In practice, due to the finite size of the unit cell, each term on the RHS of Eq.~(7) contains $\sim$30 meV per Bz due to lateral vdW interactions, which essentially cancel after the subtraction.)
The full formation energy when including the energy gained by forming an overlayer is given by the sum
of the adsorption energy and the two-dimensional (2D) assembly energy $E_{\rm form} = E_{\rm ads}+E_{\rm 2Dassembly}$.
In this paper, we approximate the assembly energy of a 2D overlayer by the pairwise sum of Bz-Bz interactions:
\bn
E_{\rm 2Dassembly}^{\rm app} \approx \half \sum_{i\ne j} E_{\rm pair} (\br_{ij})\,.
\ee
This corresponds to neglecting higher-order (multi-molecule) Bz-Bz interactions both for the van-der Waals
and indirect electronic interactions. A test calculation with a moderately dense benzene overlayer indicates
that changes in the vdW-DF result can alter the formation energy by a few tens of meV.
For indirect electronic interactions at large separations, the contribution of trio
(3-adatom, non-pairwise)  interactions decays only slightly faster than
pair interactions and should be, very roughly, 1/4 the magnitude.\cite{theo:Surf3,NEWREF}
 While such higher-order interactions are needed to achieve high accuracy,   they are not essential for our analysis.
We further argue that to a good approximation elastic effects and surface corrugation can be neglected in
computing $E_{\rm form}$. We thus calculate the direct component of the Bz-Bz interaction in isolation from
the substrate: $E^{\rm vdWDF}_{{\rm pair},h\rightarrow \infty}$. Again this energy is the difference between nearby
molecules and widely separated ones: $E^{\rm vdWDF}_{\rm pair}(d) =
E^{\rm vdWDF}(d)-E^{\rm vdWDF}(d\rightarrow \infty)$.
The interactions between molecules mediated by the surface, formally included in $E_0$, is reintroduced
by including the leading-order asymptotic part of the surface-state-mediated
indirect interaction as discussed in the previous section:
\bn
E_{\rm pair} \approx E^{\rm vdWDF}_{{\rm pair},h\rightarrow \infty} + E^{\rm ans}_{\rm pair}\,.
\ee
\noindent The asymptotic expression $E^{\rm ans}_{\rm pair }$ is expected to provide a good approximation to the full pair indirect interaction in the regime where the latter dominates the direct vdW interaction.

\subsection{Choice of unit cell, computational details}

The self-consistent charge density is found using the plane-wave and
ultrasoft-pseudopotentials code \verb DACAPO \ \cite{DACAPO} with GGA in the
PBE flavor.  For the physisorption study on a 3$\times 3$ slab unit cell
with 6 copper layers, we use a Monkhorst-Pack (3,3,1) {\bf k}-point sampling.
The study of direct vdW-interaction between Bz pairs in the dilute case was done in a
very large unit retaining only the $\Gamma$ point.  The plane-wave energy cutoff was
set higher than 500eV to ensure that the spatial gridding was consistently better than 0.14~{\AA}.


We place a Bz molecule on a 3$\times 3$ Cu(111) supercell
with a slab containing about 6 copper layers. The choice of in-surface dimensions corresponds to
a Bz-to-Bz separation of $a_{\rm cell} =\hbox{7.74 \AA}$.  Our system is large enough to extract the
essential physics but not so large as to preclude doing so in a timely fashion.  Our vdW-DF calculations are based on a
periodic unit cell and the choice of supercell does leave a small (but fixed) residual molecule
coupling arising from both direct inter-molecular vdW interactions and from substrate-mediated
interactions. The vdW-DF study of the direct vdW-interaction between Bz pairs indicates that the
effect of direct vdW bonding is small (3 times $\sim\hbox{10 meV}$); this energy is assumed to cancel in the subtraction in Eq.~(7).  However, the long-ranged
(and oscillatory) nature of the indirect electronic interaction motivates a separate
discussion as do the convergence of physisorption energies with slab thickness.

We note that the effects of indirect electronic adsorbate-adsorbate interaction will be contained
in very-large-scale DFT calculations\cite{bfDFT:Surf} and that vdW-DF would formally contain such
a contribution, primarily in the semi-local term $E_0$. However, our specific choice of supercell
with 6 copper layers does not allow more than a qualitative discussion of the surface state, Fig.~2,
and it is not possible to use our vdW-DF calculations for a quantitative analysis. Instead we
note, based on an assumption of completely absorbing surface-state scattering in the asymptotic limit, that the magnitude
of the fixed contribution from indirect electronic interactions (arising due to our use of
an implicit periodicity in our vdW-DF unit cell) can be approximated as 3$\times$2=6 meV
(according to Eq.~(5), which is not fully valid at this not-quite-asymptotic separation).
This value is much smaller than the calculated vdW-DF value for the Bz physisorption energies and we
choose not to correct for the fixed effect of implicit indirect electronic interaction in presenting
our vdW-DF results for the isolated-molecule adsorption. It is safe to assume that these inter-unit-cell contributions will not modify the vdW-DF results of extremely low diffusion barriers and
essentially corrugation-free dynamics of individual Bz adsorbates.

A partially related issue is convergence of vdW-DF energy differences with slab thickness. We
find the variation of the physisorption energy with adsorption site to be very small, a finding
that motivates us to enhance the accuracy in the vdW-DF calculations.  To test convergence with
respect to slab thickness, we compared the results of vdW-DF calculations for two different adsorbate
geometries.  Specifically, we have determined how many Cu layers $n$ are needed to accurately and
reproducibly predict energy differences for adsorption configurations
FCC-I and FCC-\R: $\Delta E^{\rm vdWDF}_{n}= E^{\rm vdWDF}_{\mathrm{FCC-I},n}-E^{\rm vdWDF}_{\mathrm{FCC-R30},n}$.
Here, $\Delta E^{\rm vdWDF}_{n}$ changed from 10$\%$ when comparing 4 and 5 layers to 2$\%$ when comparing
6 and 7 layers. We thus conclude that 6 copper layers give satisfactory convergence. The somewhat slow
convergence might be related to the need for several Cu layers before a clear signature of the Shockley
surface begins to emerge.

\section{Single molecular adsorption and diffusion}

\begin{figure}[ht]
\centering
\includegraphics[width=10cm]{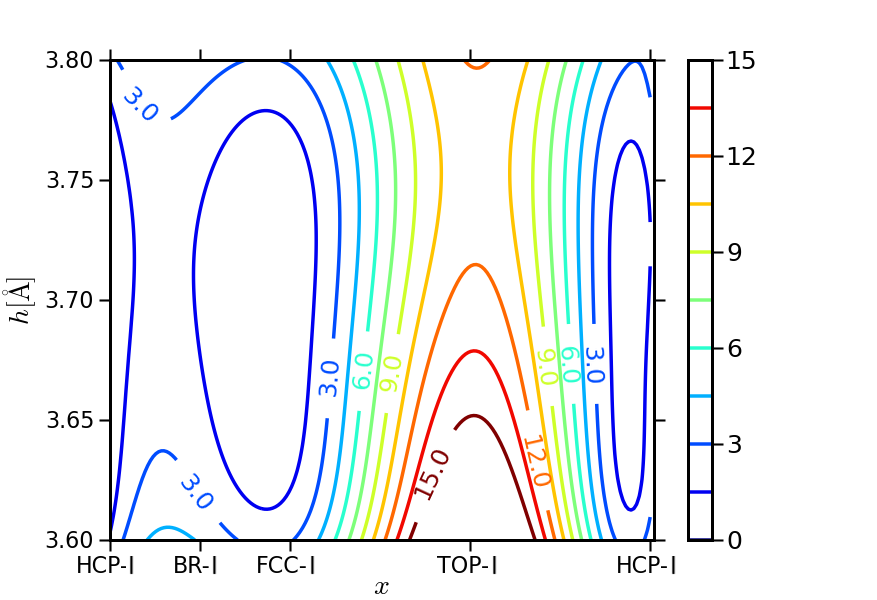}
\includegraphics[width=8cm]{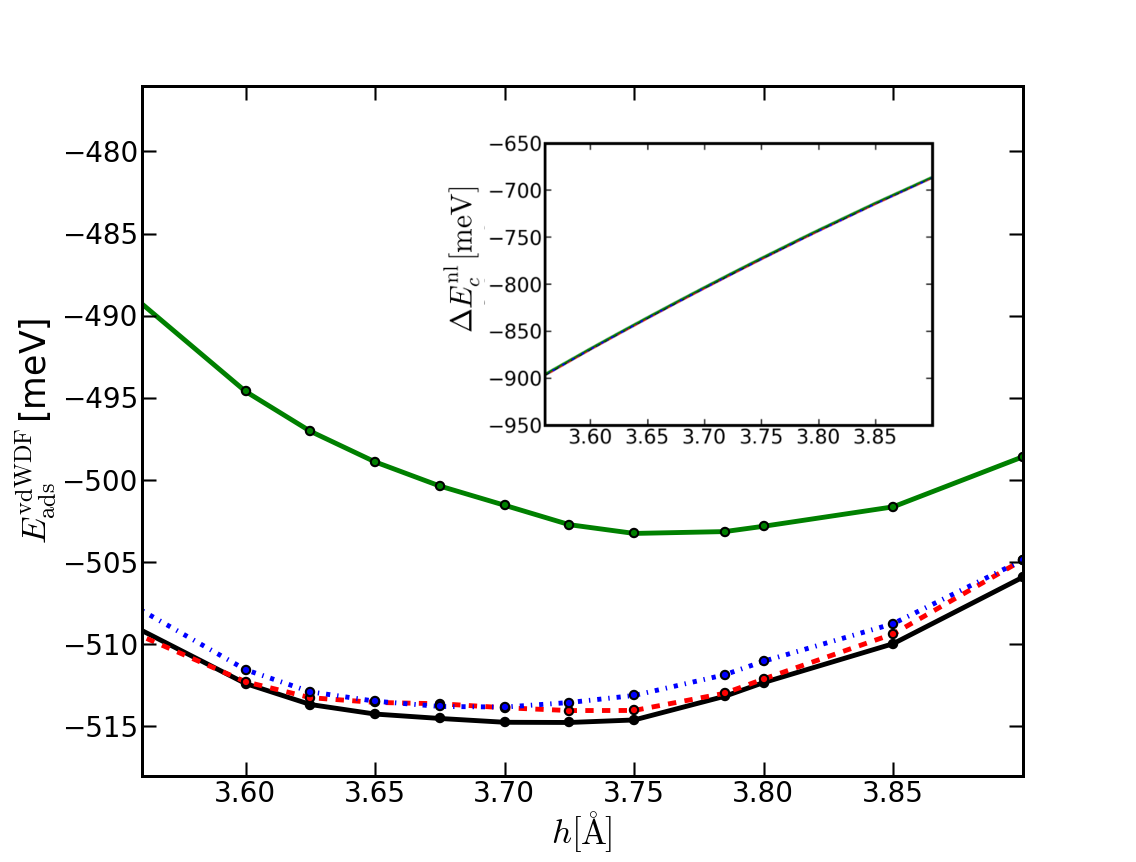}
\caption{[Color online] Potential-energy curves for Bz on Cu(111), for computing diffusion barrier. For the contour plot shown in {\it the upper panel}, the molecule is allowed to change its separation from the surface at each position. The vertical axis gives the separation from the surface and the horizontal axis shows the translation along the arrow indicated in Fig.~\ref{fig:CuLoc}. The bar to the right shows the energy scale in meV for the equipotential lines. This plot was smoothed using spline-interpolation. {\it The lower panel} shows the potential energy curves for Bz adsorption in the four high-symmetry sites with I orientation. The upper lower (upper) solid curve is for the FCC-I (TOP-I) site, the dashed (dotted) is for the HCP-I (BR-I) site. The insert in the lower panel shows the nonlocal part of the potential energy curves, which are very similar, illustrating that the energy difference between the sites stems primarily from $E_0[n]$.}
\label{fig:CuBarrier}
\end{figure}

\subsection{Physisorption energy}

Fig.~\ref{fig:CuBarrier} displays the potential energy landscape, found by performing a potential
energy search along an axis of high-symmetry sites, as shown in Fig.~\ref{fig:CuLoc}, for both BZ orientations
(configurations) I and R30$^\circ$, as defined in the caption of that figure. We allow the molecule to adjust its height at each step, but do not allow further
substrate relaxation. The spline-interpolated result for the I configurations are shown in the upper panel,
while the lower panel shows the binding curve for four of the high-symmetry sites. All I configurations have a
lower energy than all \R configurations, and the line between HCP-I, BR-I and FCC-I sites is found to be almost
degenerate, with a predicted adsorption energy of 0.51 eV, where FCC-I is slightly more favorable than the
two other sites.\cite{Wigles}

The good agreement with the experimental binding energy 0.6 eV\cite{Stevens:exp,Woll:exp} further demonstrates that the use of the vdW-DF offers a viable way to describe the binding between organic molecules and noble metal surfaces, even in the presence of a metallic surface state. The accuracy of using vdW-DF for this system is similar to its performance in general. The accuracy is also close to that found for n-butane on Cu(111).\cite{Review:vdW}

Our results clearly demonstrate the non-local nature of the binding. The binding separation of
3.75~{\AA} is a typical separation distance for physisorption; furthermore, with a Bader
analysis\cite{Bader:main,Bader:alg} we found vanishing charge transfer between the surface and the Bz molecule.

\begin{table}[t]
\begin{ruledtabular}
\caption{Interaction energies for adsorption of benzene (C$_6$H$_6$) at various positions on Cu(111) in order of decreasing stability, with vdW-DF but frozen substrate, with vdW-DF and relaxation, and without vdW-DF [from Ref.~\onlinecite{{Ford:BenzeneCuOld}}].}
\label{tab:ads}
\begin{tabular}{cccc}
Site &   $ E^{\mathrm{vdW-DF}}_{\mathrm{ads}}$[meV]&   $ E^{\mathrm{vdW-DF}}_{\mathrm{relax\, ads.} }$ [meV]& $  E^{\mathrm{GGA}}_{\mathrm{ads}}$[meV]   \\
\hline
FCC-I  & -515 & -548 &-23  \\
HCP-I & -514  & -548   &-26 \\
BR-I & -514  &  -547  &-16  \\
TOP-I & -503 & -533   &44 \\
HCP-\R & -499  & -533  &-18 \\
FCC-\R  & -498 & -527  & -17 \\
BR-\R & -497  & -529   &-21   \\
TOP-\R & -490 & -520   &10 \\
\end{tabular}
\end{ruledtabular}
\end{table}

In the results presented so far, we use a frozen substrate, thereby neglecting elastic effects. To assess the importance of these effects, we perform for every high-symmetry site a set of separate calculations, in which we let the top 4 layers of the surface relax and fix the coordinates of the benzene atoms in positions ranging from 0.1 above to 0.1 \AA\ below the optimal separations to the unrelaxed surfaces. We use PBE-GGA for this relaxation, while in the post-processing phase, the relaxation energy is estimated from vdW-DF.

Table~\ref{tab:ads} give the adsorption energy of the eight high-symmetry sites, for a frozen substrate, a substrate allowed to relaxed and for the purely PW91-GGA calculation reported in Ref.~\onlinecite{{Ford:BenzeneCuOld}}.
It shows that allowing relaxation lowers the adsorption energy within vdW-DF by between 29 and 34 meV, getting
even closer to the experimental value of  0.6 eV. Allowing relaxation gives only minor corrections to the relative
favorability of the different adsorption sites.  Note, though, that the calculations with relaxation are more
sensitive to the shifting from the PBE to the revPBE flavor of exchange in the post-processing phase, and
so are
converged only at the 2-3 meV level rather than the sub-meV convergence of the calculations without
adsorption-induced relaxation.  Since the better convergence gives us greater confidence in the
computations with a frozen substrate, we base our investigation of the surface-corrugation results on them.

Finally, Table~\ref{tab:ads} shows dramatically the need for including vdW interactions to accurately describe this adsorption system.  Calculations which do not include vdW [in this case\cite{Ford:BenzeneCuOld}, but use PW91-GGA done with the Vienna Ab-initio Simulation Package (VASP)] underestimate the adsorption energy by a factor of around 20.  For the two TOP configurations, the benzene is not even bound.  The near degeneracy of the three lowest configurations is lost, and the general ordering is considerably different.  The theme of the overall favoring of I-configurations is absent.  Thus, such calculations are inadequate for semiquantitative and perhaps even qualitative purposes.

\subsection{Surface corrugation, barrier-less diffusion on a honeycomb lattice}


Fig.~\ref{fig:CuBarrier} presents our vdW-DF results for diffusion barriers and
shows that Bz adsorbates are expected to optimize their positions relative to
their mutual interactions, with the substrate corrugation potential playing a relatively
minor role.  Table~\ref{tab:ads} presents a overview of the surface corrugation as
calculated in vdW-DF and GGA. We note that the surface corrugation (and hence diffusion
barriers) is at least an order of magnitude larger in GGA but we do not believe that the
GGA finding, top sites being unstable towards desorption, is physical.  The vdW-DF
calculations predict a much smaller corrugation: a low-temperature preference
for placing Bz adsorbates around the nearly-degenerate FCC-I, HCP-I, and BR-I
sites with a relative cost of around 11 meV for placing Bz on a TOP-I site.
Our vdW-DF description may slightly underestimate the corrugation because
it is likely to weakly overestimate the physisorption height $h$. The TOP-I
versus FCC-I/HCP-I/BR-I energy difference is in any case sufficient to affect the
expected equilibrium distribution at 77 K.  Our vdW-DF results shows that the
Bz adsorbates at low temperatures will concentrate in a honeycomb pattern
(physisorption valley) formed by linking the FCC-I, HCP-I, and BR-I sites.

At the same time, we find that the Bz adsorbates effectively experience no diffusion
barriers and are free to respond to their mutual interaction.  All of the diffusion
barriers are tiny compared to those for chemisorbed atoms; for a Cu(111) adatom the
experimental and theory results for the diffusion barrier is 40
meV.\cite{diffCu:Stoltze, exp:Surf, Cu111SiteDeterm,diffCu:Marincia}
Even for a hypothetical diffusion path, HCP-I $\rightarrow$ TOP-I $\rightarrow$ FCC-I,
the peak of the barrier (11 meV) is small enough to allow a diffusion rate larger than
1 s$^{-1}$ at $T > \hbox{4 K}$.  The diffusion barrier for motion within the honeycomb
pattern, along the path HCP-I $\rightarrow$  BR-I $\rightarrow$ FCC-I, is less than 1 meV
and we stress that the Bz adsorbates can move continuously across the surface,
using this bend-like type of diffusion. Again, our vdW-DF calculations may slightly
underestimate the diffusion barriers but it is clear that the Bz adsorbates are
free to respond to their mutual interactions on the Cu(111) surface.

Despite such low barriers, there are many reports of lattice-gas phases of physisorbed atoms at submonolayer coverages.\cite{BCZ} For example, while there is an incommensurate phase of Kr on graphite, there is also a widely-studied commensurate phase ($\surd 3\times \surd 3) R30^\circ$ with one in three sites occupied on the hexagonal grid formed by the centers of the honeycombs.\cite{KrGr}  The theoretical analyses\cite{OB} of the disordering phase transition involves a lattice-gas model appropriate to adsorption on a grid of sites.  Likewise, we find such behavior below for the two condensed phases of Bz on Cu(111).

\subsection{Rotational barriers}

Having established that the benzene can move almost freely on the surface of Cu(111), we investigate
whether it can also rotate freely. The symmetry of the I and \R configurations and the monotonic increase
in the barrier profile from HCP-I/FCC-I to TOP-I allow us to assume that the peak of the rotation barrier is
given simply by the energy difference between orientation I and orientation \R.

\begin{table}[t]
\begin{ruledtabular}
\caption{Binding difference between \R-sites and I-sites, which also give the rotational barriers. The second column show the results for a frozen substrate, while the third column gives the results when relaxation is included.}
\label{tab:rot}
\begin{tabular}{cccc}
Site &   $\Delta E^{\mathrm{vdWDF}}_{\mathrm{ads}}$ [meV] & $\Delta E^{\mathrm{vdWDF}}_{\mathrm{relax \,ads}}$[meV]  \\
\hline
HCP & 15 & 15\\
BR &  17 & 18 \\
FCC  & 17& 21 \\
TOP & 13& 13
\end{tabular}
\end{ruledtabular}
\end{table}

Table~\ref{tab:rot} compares the energy differences between Bz adsorbates in orientation \R and I, which in turn determine
the rotational barriers. The vdW-DF results show that individual molecules can easily rotate, even though the associated
activation energy is slightly higher than for diffusion.

\section{Surface assembly}
\label{s:discuss}

The vdW-DF findings of small diffusion barrier and lack of expected surface corrugation for individual Bz adsorbates
support the picture of an essentially free 2D gas of Bz molecules in the dilute regime. This is what was found on Cu
terraces at $T=77\ \mathrm{K}$.\cite{freely-Science:Weiss,freely-Surface:Weiss} Further, the finding that the location
of Bz adsorbate to some extent decouple from the underlying copper lattice, leaves the stage open for other phenomena to
dominate the interplay between molecules at higher coverages and determine the 2D surface assembly of the Bz adsorbates.

Dougherty {et al.}\cite{Dan:exp} observed the coexistence of a high density ($2.54 \mathrm{nm}^{-2}$) phase
C2 and a low density ($1.10 \mathrm{nm}^{-2}$) phase C1, both forming
hexagonal patterns. We seek insight and provide an analysis in which we first calculate the
direct and indirect Bz-Bz interactions and then combine these into an effective interaction potential.
We finally use this interaction potential to characterize the high-density phase and discuss the possibility
of surface-state-mediated forces being responsible for the structure of the low density phase.

Table \ref{tab:car} summarizes the properties of the three Bz-on-Cu(111) regimes. These results can be compared with measurements to test the accuracy and
predictive power of our vdW-DF calculations and the analysis of the indirect electronic interaction mediated by the Cu(111) surface state.\cite{Harris,theo:Surf3}
\begin{table}[t]
\begin{ruledtabular}
\caption{Energy and structure of the different phases. The formation energies (calculated from the expression for $E_{\rm form}$ after Eq. (7)) are listed both as obtained for a frozen
surface and (in parentheses) in the presence of adsorbate-induced surface relaxations.  The rotational barriers are
given by the site-orientation dependency of the benzene molecule.  For C1 and C2 data the molecules are assumed to be in the
I orientation, with the effect of indirect electronic interactions calculated for A=0.25 and $\Theta=\pi/2$
(corresponding to a choice of completely absorbing scattering). The square brackets give the values with $A=0$. The rotational excitation is calculated under the assumption that the I-\R interaction is limited to
neighboring molecules.}
\begin{tabular}{lcccc}
 &  C2   & C1 &  Dilute  \\
\hline
Formation energy [meV] &  618 (--)&  539 (572)   & 515 (548) \\
Exp't adsorp'n energy [meV] & - &    -  &  600$^a$   \smallskip \\
Lattice parameter [\AA]  &  6.9  & 10.2   & -  \\
Exp't lattice parameter [\AA]  & 6.8 $^b$ & 10.2$^b$  & - \smallskip  \\
Rotation barrier [meV] & -   & $ \sim \! 17$  & 17 \\
Rotation excitation [meV]    &   $\sim 9$ & - & -\\
2D-assembly energy [meV] & $ \sim \! 79$ [97] &  $ \sim \! 24$ &  - \\

\end{tabular}
$^a$ Based on Ref.~\onlinecite{Woll:exp}.
$^b$ Based on Ref.~\onlinecite{Dan:exp}.
\label{tab:car}
\end{ruledtabular}
\end{table}

\subsection{Results for pair interactions}

\begin{figure}[t]
\centering
\includegraphics[width=8cm]{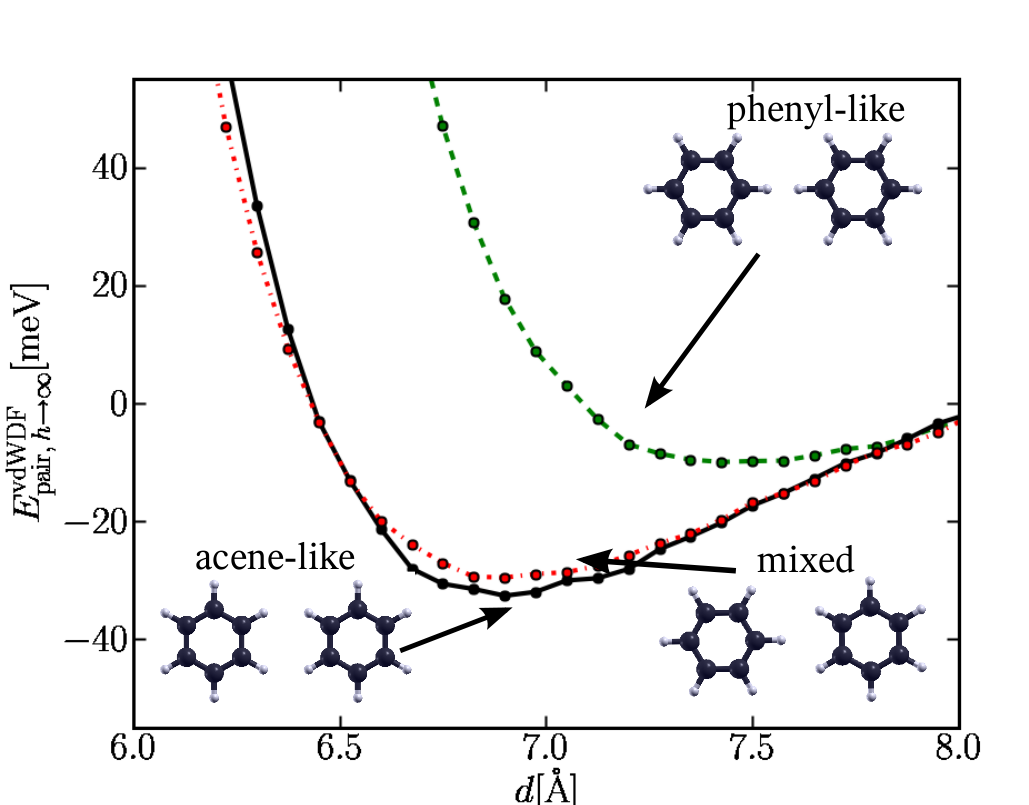}
\caption{[Color online]
Binding curve of two Bz molecules in different high-symmetry configurations identified by the set of inserts and calculated
in vdW-DF in isolation from the surface. The two Bz's lies entirely
within a common plane and the abscissa gives the distance between the center of two Bz's.  We label the configurations shown in the lower left, lower right and upper right corner as acene-like(A), mixed(M) and phenyl-like(P). The most favorable A configuration (solid curve)
has an optimal binding distance of 6.9 \AA, and corresponding
vdW-DF binding energy is 32 meV.
}
\label{fig:BenzDimer}
\end{figure}

There are two important forces between the molecules for the interacting system formed
when multiple Bz adsorbates assemble on Cu(111).  First is the direct vdW interaction between
the Bz molecules; this interaction is medium range.  Second is the indirect electronic
Bz-Bz interaction, mediated by the surface-state
electrons, which is long-range and oscillatory.\cite{exp:Surf, theo:Surf1, theo:Surf2} In this subsection, we address the
influence of these effects for pair interactions and construct an effective pair potential
which can be used to discuss island formation and phase condensation of Bz on Cu(111).\cite{Dan:exp}

Fig.~\ref{fig:BenzDimer} shows the vdW-DF binding curve of
Bz pairs, $E^{\rm vdWDF}_{{\rm pair},h\rightarrow \infty}(d)$
for different high-symmetry configurations identified in the set of
inserts and calculated in vdW-DF ignoring contact with the surface.
The two Bz molecules lies entirely within a common plane. For lack of
better notation in the litterature, we label them acene-like (A), phenyl-like (P) and mixed (M). In the
A (P) configuration the displacement vector from one Bz to the other goes
perpendicularly through an edge (a vertex) of both carbon hexagons.
In the M configuration the
displacement vector goes through a vertex of one Bz and an edge of the other. The binding energy is largest
for the acene-like configuration, closely followed by the mixed, with the minimum of the bonding curve at
$-32$ meV. In the P configurations bonding is much weaker and at a larger separation. The direct vdW binding clearly has a
larger magnitude than the barrier profile for the adsorbed molecules. Thus, the Bz-Bz pair interactions should dominate over surface
corrugation effects at smaller separations ($d$ smaller than $\sim \hbox{8 {\AA}}$). Summing the contributions of pairwise vdW bindings in A configurations, we estimate the full vdW-assembly energy of an isolated sheet of Bz molecules in a hexagonal lattice to be 97 meV.

\begin{figure}[t]
\centering
\includegraphics[width=8cm]{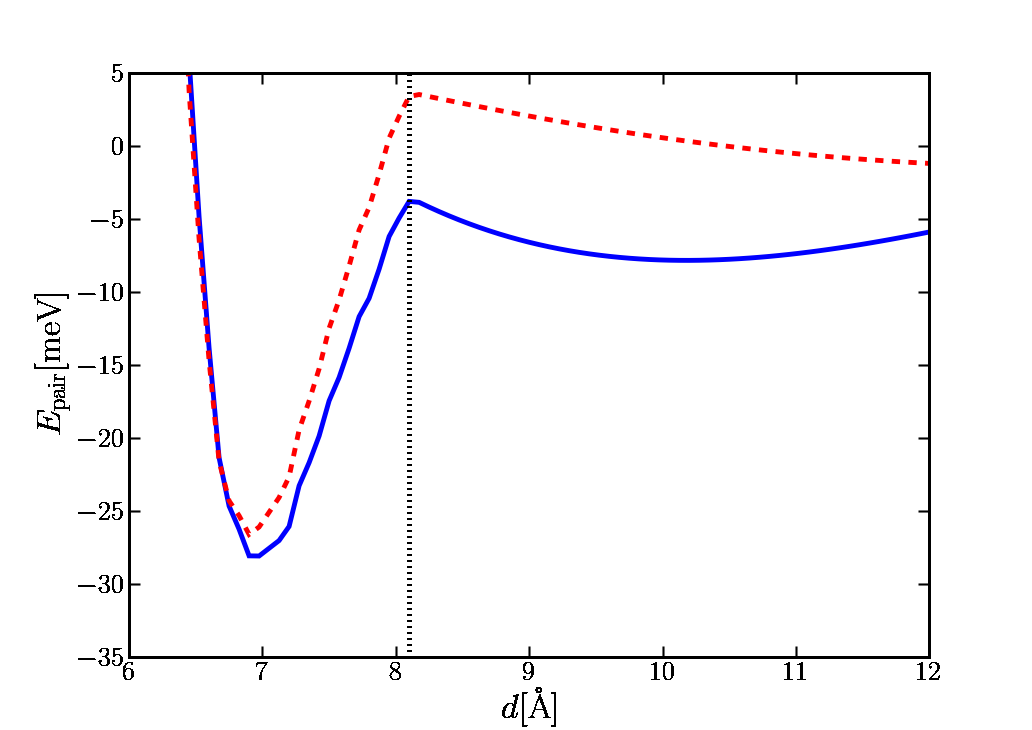}
\caption{[Color online] Effective binding potential of a Bz pair from the combinations of direct vdW and indirect electronic
(surface-state mediated) Bz-Bz interactions, assuming favorable orientations of Bz molecules. The upper dashed curve shows
the results for the values of $A=0.08$ and $\Theta=0.3\pi$, while the lower solid curve shows the results for $A=0.25$
and $\Theta=\pi/2$. For $d > 8.1$ \AA \ (indicated by the dotted black line), the vdW interaction is negligible,
and the displayed interaction is given by the formal Harris-functional evaluation, Eq.~(\ref{eq:pair}).}
\label{fig:CombinedBarrier}
\end{figure}

On the Cu(111) substrate, the I-configurations are optimal for single-molecular adsorption and the
A configurations are optimal for pairs. It is thus relevant to
discuss an effective adsorbate-pair interaction $E_\mathrm{pair}$ with
implicit assumptions of favorable molecular orientations (with both Bz
molecules in surface orientation I and with the direct vdW interaction
specified by the A configuration). This effective (pair) potential
determines adsorbate organization into a well-ordered triangular lattice
structures at very low temperatures.

Fig.~\ref{fig:CombinedBarrier} shows our result for the effective
(favorable-condition) potential $E_\mathrm{pair}$ with a binding curve
that includes the asymptotic surface-mediated interactions given by
Eq.~(\ref{eq:pair}). Since we lack explicit information about the
detailed nature of surface-state scattering by Bz adsorbate, we must
make assumptions for it by analogy with the case of atomic adsorption on
Cu(111).\cite{exp:Surf}


We give the results for the case of a completely absorbing scatterer with $A=0.25$ and $\Theta=\pi/2$,
and, for comparison, the values fitted to the experimental scattering of copper adsorbates \cite{exp:Surf} with
$A=0.08$ and $\Theta=0.3\pi$. The black dotted vertical line in the figure separates the region in which
van der Waals forces are important from the regions where the surface-mediated interactions dominate.
For the latter region only the indirect-electronic interactions are included. For $A=0.25$, the curve has two minima,
the first stemming from direct intermolecular interactions, and the second, weaker one stemming from the
sinusoidal form of the contribution~(\ref{eq:pair}). For $A=0.08$, the indirect-electronic interaction only
creates a second, very weak minimum (1.5 meV) at 14 \AA, but do still create an energy barrier for the
adsorbates trying to move from the C1 phase and assemble into the vdW-bound C2 phase.

\subsection{Van der Waals lattice (C2)}
\begin{figure}[t]
\centering
\includegraphics[width=8cm]{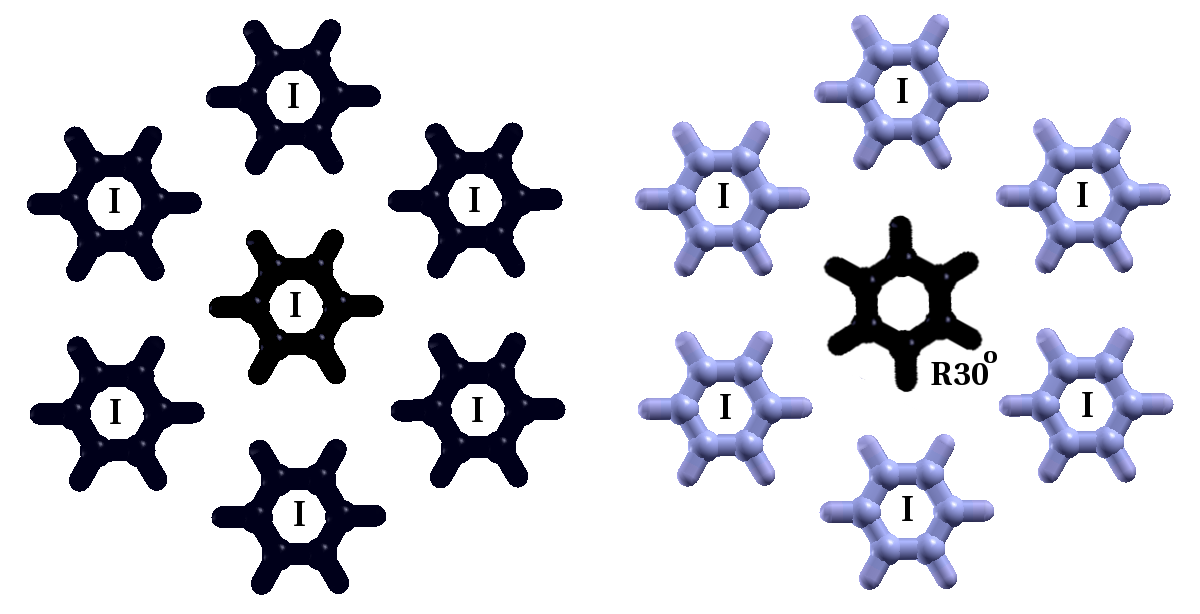}
\caption{[Color online] Schematic illustration of a rotational excitation in the hexagonal phase C2 and of
the interlocking effects which (at low temperatures) will prevent neighboring adsorbates from rotating
until the original rotational excitation relaxes. The figure shows two near-degenerate configurations of
benzene on Cu(111), in the left panel all molecules are in the ground-state packing with all adsorbates in the
I-orientation, while in the right picture one of molecules has switched to the \R-orientation.
This switch corresponds to six adsorbate pairs changing to the M configuration.
For pairs, this switch corresponds to six of them changing to the M configurations.
While this rotational excitation does not cost much energy per se, it does block the rotation of all the neighboring molecules (marked in light blue [gray in print]) until the center molecules relaxes back to orientation I.}
\label{fig:CombPhase}
\end{figure}

The optimal pair separation of 6.9 \AA , as shown in Figs.~\ref{fig:BenzDimer} and
\ref{fig:CombinedBarrier}, illustrates that it can be favorable for islands to form at benzene
separations somewhat shorter than that of the 3$\times$3 unit cell: $3a \approx 7.74$ {\AA}.
The primary limitation for the density of these islands is the kinetic-energy repulsion between benzene molecules.
Assuming that the molecules lie flat on the copper surface and form a hexagonal pattern with pairs of molecules in A configurations, we estimate a density of 2.42 nm$^{-2}$, very close to the experimentally determined density.

Fig.~\ref{fig:CombPhase} illustrates a curious consequence for the adsorbate dynamics in phase C2. We include it
as the phenomenon is a consequence of the proposed interpretation of phase C2 as a planar hexagonal pattern of Bz
adsorbates, primarily organized in orientation I, with pairs of Bz's in the A configuration and assembled by direct vdW bonding; an experimental observation of this consequence would constitute a specific test of the proposed interpretation of phase C2 as a vdW-bound
assembly.  Based on vdW-DF calculations of the direct bonding, we find that the hexagonal
pattern will permit a low-energy excitation with a rotation of a single Bz adsorbate from orientation I
to \R. This switch corresponds to six Bz pair configurations changing from
A to M.
However, as illustrated in the right panel, such a rotational excitation will (at low temperature) cause a blocking of rotations for all neighboring Bz adsorbates.
This effect is due to the strong repulsion at the typical C2 separation for pairs of Bz's in the P configuration.

\subsection{Friedel lattice (C1)}

The asymptotic form of the pairwise indirect-electronic interaction, shown in Fig.~\ref{fig:CombinedBarrier}, predicts
a second minimum of only 8 meV, comparable to the calculated magnitude of surface-corrugation effects.
However, in forming a lattice, there are multiple pairs contributing, effectively increasing the impact of this
effect by a factor of 3. Again, assuming a hexagonal lattice and using $A=0.25$, we get the experimental minimum
of $a=10.2$ \AA \ and a binding energy of $3\times8 {\rm meV}=24 {\rm meV}$.

The overlayer lattice parameter in phase C1 might be also influenced by the surface corrugation, since the overlayer is
essentially commensurate with the 4$\times$4 copper unit cell, with a lattice parameter of 10.21 \AA. The softness
of the binding curves of Fig.~\ref{fig:CombinedBarrier} supports this idea. The diffusion barriers found
in this study, however, suggests that it is hard to account for the stability of the C1-phase without including the
surface-mediated interactions. Even if the scattering properties of the benzene adsorbate are closer to
those of copper adsorbates ($A=0.08,\, \Theta=0.3\pi$), and therefore does not determine the overlayer-lattice
parameter, the resulting barrier should contribute to the stability of this phase.


\section{Conclusions}

The low barrier to diffusion implies that benzene adsorbates can respond to weak interactions. The low barriers could,
more generally, allow acene adsorbates to self-organize in response to either (or both) direct vdW interactions or
(and) indirect surface-state-mediated interactions.  There are several conditions for
this to occur.  The adsorption of the acene must be sufficiently weak that the process does not destroy
the metallic surface state yet strong enough that the interaction it mediates can account
for the regularity of the honeycomb network.  For benzene molecules we have shown that these conditions are
present and that the hexagonal-lattice ordering found in the C1 and C2 phases can be interpreted as
arising from substrate-mediated and vdW binding, respectively.

\section{Acknowledgments}

The authors thank SNIC (Swedish National Infrastructure for Computing)
for access and for KB's participation in the graduate school NGSSC.
Work at U. of Maryland was supported primarily by NSF Grant No.\ CHE 07-50334,
with secondary support from NSF-MRSEC Grant No.\ DMR 05-20471 and from the Computational Materials Science Network of DOE, and ancillary support
from the Center for Nanophysics and Advanced Materials (CNAM). Work at Chalmers was supported by the Swedish
Research Council (Vetenskapsr{\aa}det VR) grant 621-2008-4346.
The authors are grateful for exchanges with Daniel B. Dougherty.
The authors thank {\O}yvind Borck for use of his code for Bader analysis.
TLE acknowledges extensive collaboration and helpful interactions with
Ludwig Bartels and Kwangmoo Kim.


\begin{thebibliography}{5}
\bibitem{Wigles} As an interesting aside, we note that the vdW-DF differences in Bz-physisorption site-preference
shown in Fig.~\ref{fig:CuBarrier} originates almost entirely from difference in $E_0$ (from differences in
kinetic-energy repulsion) while the nonlocal-correlation energies differ by at most 2-3 meV between the sites
FCC-I, HCP-I, BR-I, and TOP-I.  Moreover, the fluctuations (numerical noise) in the evaluation of the
nonlocal-correlation energy $E_c^{\mathrm nl}$ is very small, converged to the sub-meV level.
Rather than being an issue with the $E_c^{\mathrm nl}$ evaluation, we have traced the origin of the small
wiggles in the vdW-DF binding curve in the lower panel of Fig.~\ref{fig:CuBarrier} to $E_0$, specifically,
the post-processing adjustment of the exchange from PBE to revPBE.
\end{thebibliography}
\end{document}